%% file: main.tex
\documentclass[10pt,conference]{IEEEtran}
\usepackage{fullpage,url,times}
\usepackage{graphics,graphicx,subfigure,epsfig,multicol}
\usepackage{amssymb,amsmath,amsbsy, amstext, algorithm,algorithmic}
\usepackage{color}

\newtheorem{theorem}{\bf Theorem}[section]
\newtheorem{definition}{\bf Definition}
\newtheorem{lemma}[theorem]{\bf Lemma}

\begin{document}

\title{Stateless and Delivery Guaranteed Geometric Routing on Virtual Coordinate System}

\author{Ke Liu and Nael Abu-Ghazaleh \\
CS Dept., SUNY Binghamton \\
\url{{kliu,nael}@cs.binghamton.edu}}

\maketitle

\input{abstract}

\input{introduction}

\input{related}

\input{vcdegrade}

\input{spvc}

\input{experiment}

\input{conclusion}

\bibliographystyle{plain}
\bibliography{main.bib}

\end{document}

%% file: abstract.tex
\begin{abstract}
Stateless geographic routing provides relatively good performance at a
fixed overhead, which is typically much lower than conventional
routing protocols such as AODV.  However, the performance of
geographic routing is impacted by physical voids, and localization
errors.  Accordingly, virtual coordinate systems (VCS) were proposed
as an alternative approach that is resilient to localization errors
and that naturally routes around physical voids.  However, VCS also
faces virtual anomalies, causing their performance to trail geographic
routing.  In existing VCS routing protocols, there is a lack of an
effective stateless and delivery guaranteed complementary routing
algorithm that can be used to traverse voids. Most proposed solutions
use variants of flooding or blind searching when a void is
encountered.  In this paper, we propose a spanning-path virtual
coordinate system which can be used as a complete routing algorithm or
as the complementary algorithm to greedy forwarding that is invoked
when voids are encountered.  With this approach, and for the first
time, we demonstrate a stateless and delivery guaranteed geometric
routing algorithm on VCS.  When used in conjunction with our
previously proposed aligned virtual coordinate system (AVCS), it
out-performs not only all geometric routing protocols on VCS, but also
geographic routing with accurate location information.
\end{abstract}

%% file: introduction.tex
\section{Introduction}
\label{sec:introduction}

In contrast to traditional ad hoc routing protocols such as
AODV~\cite{paper:aodv}, Geographical routing
~\cite{paper:gfg,paper:gpsr,paper:goafr+,paper:goafr++,paper:gcrp,paper:practical_gr,paper:pvfr,paper:gfg_new,paper:near},
provides attractive properties for multi-hop wireless networks.
Specifically, geographic routing operates via local interactions among
neighboring nodes and requires a fixed and limited amount of state
information that does not grow with the number of communicating nodes,
(therefore, it is called {\em stateless}). Nodes exchange location
information with their neighbors. Packets addressed to a destination
must provide its location. At every intermediate hop, the subset of
the neighbors that are closer to the destination than the current node
is called the forwarding set (FS). Routing simply forwards a packet to
a node in FS, typically the one closest to the destination.  This
process is repeated greedily until the packet reaches the
destination. Thus, interactions are localized to location exchange
with direct neighbors.

Geographical routing protocols suffer from significant problems under
realistic operation.  First, {\em voids} --intermediate nodes whose FS
relative to a destination is empty-- can cause the greedy algorithm to
fail \cite{paper:gfg,paper:gpsr,paper:bphole,paper:gcrp}. Voids
require a somewhat complex and inefficient complementary routing
algorithm (e.g., perimeter routing) that is invoked when they are
encountered.  Perimeter routing requires more information in addition
to the location of neighbors \cite{paper:practical_gr}. Moreover,
geographic routing has been shown to be sensitive to localization
errors~\cite{paper:errorgf}, especially in the perimeter routing
phase~\cite{paper:practical_gr,paper:errorfr}; such errors can cause
routing anomalies ranging from suboptimal paths to loops and failure
to deliver packets. Making geographical routing protocols practical is
extremely difficult~\cite{paper:practical_gr}.

Routing based on Virtual Coordinate Systems (VCS) has been recently
proposed~\cite{paper:vcembed,paper:lcr,paper:vcsim,paper:vcap,paper:bvr,paper:gspring,paper:wiredgr}
to address some of the shortcomings of geographic routing.  A VCS
overlays virtual coordinates on the nodes in the network based on
their network distance (typically in terms of number of hops) from
some fixed reference points; the coordinates are computed via an
initialization phase. The virtual coordinates serve in place of the
geographic location for purposes of geographic forwarding; that is, in
these algorithms the FS is the set of nodes that are closer to the
destination than the current node, based on a function that computes
distance between points in coordinate space (e.g., Cartesian distance,
or Manhattan distance).  Because it does not require precise location
information, VCS is not sensitive to localization errors. Further, it
is argued that VCS is not susceptible to conventional voids because
the coordinates are based on connectivity and not physical distance
\cite{paper:lcr}.  On the negative side, VCS may be sensitive to
collisions and or signal fading effects in the initialization phase.
Furthermore, the initialization phase requires a flood from each
reference point.  Finally, the coordinates should be refreshed
periodically if the network is dynamic.  Both geographic and
virtual coordinate routing represent instances of {\em geometric routing}.

Existing research work in geometric routing protocols concentrates on
optimizing different aspects of existing coordinate systems
\cite{paper:gpsr,paper:gfg,paper:practical_gr,paper:gfg_new,paper:lcr,paper:bvr,paper:gspring,paper:wiredgr}.
Why and how the virtual anomalies occur in VCS routing is a topic that
has not received attention.  In previous work, we categorized some of
the reasons behind VCS anomalies,
\cite{paper:hgr,paper:avcs}.  For example, we
identified and explained the disconnected VCS zone problem (where
unconnected nodes may receive the same coordinates).  We also
identified a group of anomalies that arise due to the quantization
error present in an integer VCS being overlayed over a continuous
space.  However, a systematic analysis of all the causes remains
elusive.

The first contribution of this paper is to analyze the reasons causing
the virtual anomalies systematically from the perspective of the limit
of graph connectivity. More specifically, since the connectivity of
the network's mapping graph in practice varies, with regions that are
not well connected, the uniqueness of nodes' coordinate identities
also varies, causing any VCS with a fixed number of anchors to
potentially fail to provide guaranteed delivery. For example, a
1-connected network (linear chain) does not benefit from VCS with more
than 1 virtual dimension, and meanwhile, an n-connected graph may
benefit from increasing number of virtual dimensions (anchors) beyond
n. Consequently, in practice, any VCS with an arbitrary number of
virtual dimensions may suffer from degraded connectivity and end up with multiple nodes sharing the same coordinate value in the network.

The second contribution of this paper is to propose a new VCS:
Spanning-Path virtual coordinate system (SPVCS), providing a universal
unique identity to any node in network, based on the conservative
assumption that the network is only 1-connected (if a network is
1+-connected, it is also 1-connected). Based on SPVCS, a stateless and
guaranteed loop-free geometric routing path can be constructed. We
call this routing algorithm the Spanning-Path Geometric Routing (SPGR)
algorithm.  An optimization of SPVCS is also proposed (OSPVCS), which
improves the routing performance in term of path stretch of SPGR.

The third contribution of this paper is to explore using SPGR with our
previous work, the aligned virtual coordinate system (AVCS), leading
to a stateless and delivery guaranteed geometric routing protocol with
a much better path stretch relative to other VCS and geographic
routing protocols.  Specifically, in this approach, we use the
efficient AVCS for the greedy phase of the algorithm, reserving SPGR
for the complimentary phase when an anomaly is encountered.  We call
the resulting protocol the {\em aligned greedy and spanning-path}
(AGSP) routing protocol. 

We use simulation to compare the performance of geometric routing
protocols on different coordinate systems, such as geographic
coordinate system, VCS, the aligned VCS and SPVCS. The experimental
results show that AGSP on AVCS and SPVCS outperform other other
geometric routing protocols including GPSR/GFG
\cite{paper:gpsr,paper:gfg}, LCR \cite{paper:lcr} and BVR
\cite{paper:bvr}.

The remainder of this paper is organized as follow: Section
\ref{sec:related} provides an overview of background and related
works. After analyzing the systematic reason causing virtual anomalies
with VCS in Section \ref{sec:vcdegrade}, we present the design of
Spanning-Path VCS and the routing protocol that uses it in Section
\ref{sec:spvcs}. In Section \ref{sec:experiment}, the experimental
study is presented to compare most existing geometric routing
protocols on different coordinate system. Finally, we conclude in
Section \ref{sec:conclusion}.

%% file: related.tex
\section{Background and Related Work}
\label{sec:related}
Stateful hop-count based routing protocols such as AODV
\cite{paper:aodv}, are commonly-used in Ad hoc networks.  A variant,
called Shortest Path (SP), can be used in sensor networks where data
is funneled to a few sinks: in SP, data sinks send periodic
network-wide beacons (typically using flooding). As nodes receive the
beacon, they set their next hop to be the node from which they
received the beacon with the shortest number of hops to the
sink. Thus, with a single network wide broadcast, all nodes can
construct routes to the originating node. SP generally provides the
optimal path in terms of path length.  However, it is a stateful and
reactive protocol: for each data sink, the forwarding path is needed
before data transmission can begin.  The required storage increases
with the number of destinations in the network.  Furthermore, SP is
vulnerable to mobility or other changes in the topology.

To counter these disadvantages, stateless geometric routing protocols
were proposed.  GFG~\cite{paper:gfg}, and the very similar
GPSR~\cite{paper:gpsr}, are the earliest and most widely used of this
class of protocols.  They consist of a Greedy Forwarding (GF) phase
where each node forwards packets to the neighbor that will bring the
packet closest to the destination.  Each node tracks only the location
information of its neighbors.  Based on this information, for a packet
with a given destination, a node can determine the set of neighbors
closer to the destination than itself; this set is called the {\em
forwarding set} for this destination.  GF proceeds by picking a node
from this set, typically the closest to the destination.

It is possible that GF fails, if the forwarding set is empty: a {\em
  void} is encountered.  A complementary phase of the algorithm is
  then invoked to traverse the void.  Typically, face routing or
  perimeter routing; this is an approach based planar graph theory.
  The general idea is to attempt to route around the void using a
  right hand rule that selects nodes around the perimeter of the void
  (details may be found in the original paper~\cite{paper:gpsr}). This
  approach is continued until a node closer to the destination than
  the void origin is encountered; at this stage, operation switches
  back to greedy forwarding.  However, a problem arises if the
  perimeter routing intersects itself -- there is a danger that the
  packet gets stuck in a loop.  Thus, a technique for planarizing the
  graph to avoid the use of intersecting edges is needed: {\em
  Relative Neighborhood Graph (RNG)} and {\em Gabriel Graph (GG)} are
  2 kinds of such planarization techniques.

GPSR and other geographic routing protocols are vulnerable to
localization errors.  Since GPS devices are costly, they may not be
feasible for sensor networks; often, localization algorithms are
employed that significantly increase the uncertainty in the location
estimate (e.g.,
~\cite{paper:aps,paper:hightower01,paper:robustloc-mobicom2002}). The
degree of error in the location estimate depends on the localization
mechanism (an error up to 40\% of the radio range is considered a
common case).  Both the greedy forwarding and face routing phases are
susceptible to localization
errors~~\cite{paper:errorgf,paper:errorfr}. While some approaches to
tolerate location errors have been suggested, in general, this remains
a weakness of this class of protocols.  Further, the paths constructed
by face routing are typically extremely inefficient, especially if the
network is dense.  Thus, additional routing protocols have attempted
to optimize the face routing phase of
operation~\cite{paper:gcrp,paper:bphole,paper:glider,paper:pvfr,paper:gfg_new,paper:near}.
However, most of these works optimize face routing in term of path
quality, but tend to increase the overhead and the complexity.  They
do not address the effect of location errors on the improved schemes.

Routing based on a coordinate system, rather than location, was
first proposed by Rao et al~\cite{paper:grnoloc}.  However, this
approach requires a large number of nodes to serve as virtual
coordinate anchor nodes (sufficient to form a bounding polygon
around the remaining sensors).  The drawback of having many
reference points is that forming coordinates requires a long time to
converge; the same is true for the overhead to refresh coordinates.
Instead of using the virtual coordinates directly for routing, Rao
et al use them to estimate location for use in geographic routing.
Reach-ability is an issue in this protocol as geographic location is
approximate; recall that it has been shown that both the greedy
forwarding and the face routing phases of geographic routing are
susceptible to localization errors. Similar approaches that use VCS
to aid localization have been also used by other
works~\cite{paper:gc,paper:aps}.  Essentially, these works collapse
the original VCS coordinates back into 2 geographic coordinates for
the purpose of routing.

GEM~\cite{paper:gem} proposed routing based on a virtual coordinate
system. A virtual polar coordinate space (VPCS) is used for localizing
each node in the network. A tree-style overlay is then used for
routing. Thus, GEM is not stateless. Further, GEM works only as a
localization algorithm, generally does not provide guaranteed
uniqueness of node identity based on coordinates. Since it uses the
VPCS to localize the network first, it tolerates only up to $10\%$
localization error~\cite{paper:gem}.

Caruso et al proposed the Virtual Coordinate assignment protocol
(VCap) \cite{paper:vcap}.  Several similar protocols are also proposed
\cite{paper:vcembed,paper:vcsim,paper:lcr,paper:bvr,paper:gspring,paper:hgr,paper:avcs,paper:wiredgr}.
In this approach, coordinates are constructed in an initialization
phase relative to a number of reference points. Following this
initialization phase, packets can be routed using the Greedy
Forwarding principles, replacing node location with its coordinates:
the forwarding set consists of neighbors whose coordinates are closer
(different distance functions have been proposed) to the destination
than the current node.  Caruso et al advocate the use of 3 reference
points to assign the virtual coordinates, constructing a 3-dimensional
VCS. We showed that this 3D VCS may not sufficient to map the network
effectively\cite{paper:hgr}. VCap, even with 4 coordinates performs
significantly worse than GPSR both in delivery ratio (node pair
reach-ability) and path quality. Qing et al proposed a similar
protocol to VCAP, called Logical Coordinate Routing (LCR), with 4
reference nodes located at the corners of a rectangular
area~\cite{paper:lcr}.  LCR proposes a backtracking algorithm for
traversing voids; however, it requires that each node remember every
packet that passes through it.

Rodrigo et al proposed beacon vector routing (BVR) \cite{paper:bvr},
which forms a VCS with a large number of anchors (typically 10 to 80).
BVR uses Manhattan-style distance, whereas VCAP and LCR use Euclidean
distance, to measure distance between two given coordinate points.
BVR uses such a large number of anchors to increase the possibility of
BVR routing success in the greedy mode. However, even with so many
anchors, BVR fails frequently for scenarios that we evaluated.  BVR
proposes the use of a backtracking approach upon failure to forward
packets back to the reference node closest to the destination when
greedy forwarding fails.  Once the beacon receives this packet, it
floods it towards the destination.

Few works have explicitly analyzed the reasons behind VCS coordinate
routing failures.  Moreover, existing protocols for complementary
routing are heuristic in nature, and often quite complex in terms of
their state requirement.  In previous work, we identified
quantization errors as one of the reasons for VCS anomalies, and
proposed an aligned virtual coordinate system, where each node
averages its coordinates with those of its neighbors, to reduce this
quantization error~\cite{paper:avcs}.  AVCS significantly reduces,
but does not completely eliminate, the onset of anomalies in VCS.
Leong et al proposed a similar improvement to AVCS which they call
GSpring \cite{paper:gspring}.  However, GSpring requires a dynamic
re-construction of the virtual coordinate system during routing,
leading to a long convergence time.  Moreover, their achieved
performance does not exceed that of AVCS.  Huang et al proposed
network dilation \cite{paper:dilation} to resolve similar anomalies;
Dilation requires a complicated mathematic model, and no routing
protocol has been demonstrated to capitalize on it.

Papadimitriou and Ratajczak~\cite{paper:gfconjecture} conjecture that
a greedy embedding can be always found in a 4-connected graph, which
means if a network is 4-connected, we can always find a greedy routing
algorithm to be delivery guaranteed. Furthermore, Rote and
B\'{a}n\'{a}ny proved that every planar 3-connected graph can be
embedded on the plane so that greedy routing works
\cite{paper:greedydrawings,paper:greedydrawingsFull}. However, in
reality, a fully 4-connected graph is not a common network topology.
A 3-connected planar graph is even more difficult to construct since
most existing graph planarizing algorithm requires the physical
coordinates of all nodes in network \cite{phdthesis:benleong}.

%% file: vcdegrade.tex
\section{What causes Virtual Void?}
\label{sec:vcdegrade}

In previous work \cite{paper:hgr,paper:avcs}, we
analyzed several categories of virtual anomalies. Although some
reasons for virtual anomalies were identified, some anomalies remained
unexplained. In this section, we generalize the explanation virtual
anomalies and show how this reason subsumes the explanation for the
virtual coordinate anomalies presented in our previous work.

\subsection{Dimension Degradation}

\begin{figure}[!t]
\centering
\includegraphics[width=0.35\textwidth]{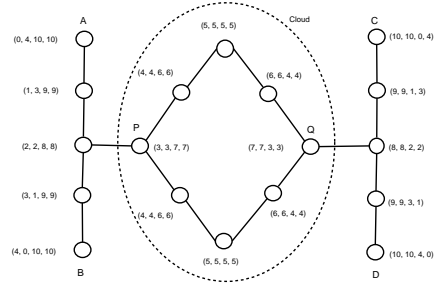}
\caption{Example: Dimension Degradation} \label{fig:dimDegradation}
\end{figure}

In virtual coordinate systems (VCS), it is desirable to minimize the
onset of anomalies so that greedy forwarding works more frequently.
The intuition behind some of the emerging VCS designs is that the
uniqueness (measured in terms of percentage unique node labels) of
the naming algorithm is positively related to the number of
dimensions in a VCS. This intuition is based on an implicit
assumption: the network is {\em N-connected } for an {\em
N-dimensional} VCS and the uniqueness continues to increase as the
number of dimensions increases. Figure \ref{fig:dimDegradation}
shows a network with a 4-dimensional VCS, by setting 4 anchor nodes
at node A, B, C and D. The network is just 2-connected. And the
highest degree of any vertex is only 3. In this case, the
4-dimensional VCS does not increase the naming uniqueness from a
2-dimensional VCS with anchor nodes as A and C (or B and D). The
nodes in the middle {\em Cloud} can mostly find another node with
the same identity, either in a 2-dimensional or 4-dimensional VCS,
except the {\em vertex cut} nodes P and Q. Continuing to increase
the number of dimensions of the VCS would not help the naming
uniqueness, if the additional anchors locate outside the cloud. We
call the highest number of VCS dimensions can be used to increase
naming uniqueness the {\em dimension}; when this dimension is less
than $N$ we refer to this phenomena as {\em dimension degradation}.

\begin{definition}
Given a graph $G(V, E)$, a {\bf component} of it is a graph $G'(V',
E')$ where $V' \subseteq V$, $E' \subseteq E$ and $|V'| \geq 2$.
\end{definition}
\begin{definition}
A {\bf node cut} (or a vertex cut) of a component $C(V', E')$
is a set of nodes $V_c \subseteq V'$ where removing $V_c$ will
disconnect the rest of $C$ from $G-C$ or $|V'| = |V_c|$.
\end{definition}
\begin{definition}
The {\bf connectivity} of a graph $G$ is the minimum size node cut. A
graph is {\bf $k$-connected} if its connectivity is at least $k$.
\end{definition}
\begin{definition}
A {\bf determinant component} of a network with VCS, is a component of
the mapping graph of this network, containing one or more VCS anchors.
An {\bf indeterminate component} is a component which is not
determinant.
\end{definition}
\begin{definition}
The virtual coordinate {\bf uniqueness degree} $U_d$ of a VCS is the
number of unique virtual coordinates of all nodes.
\end{definition}
\begin{definition}
A {\bf dimensional degradation} $D_d$ is the maximal number of
dimensions of a network which can increase its $U_d$.
\end{definition}
For example, if the $U_d$ of a n-dimensional VCS on a network is
$x$, and the $U_d$ of a (n+1)-dimensional VCS is also $x$, we say
the $D_d$ of this network is $n$.

\begin{theorem}
The $D_d$ of a {\em $1$-connected} graph is 1.
\end{theorem}
\begin{proof}
Suppose a graph (of some network) $G(V, E)$ is $1$-connected, and
the vertex cut is $v \in V'$, where $C(V', E')$ is an {\bf
indeterminate} component of $G$ and $|V'| > 1$. In 1-dimensional
virtual coordinate system, the virtual coordinate value of $v$ is
$P^{(1)}=(p_1)$. For $\forall v_i \in V'$, its network distance to
$v$ is $d_i$ (in number of hops). We can directly infer the virtual
coordinate value of $v_i$ is $P_i^{(1)} = (p_1 + d_i)$. We need to
prove that $\forall v_x, v_y \in V'$ if $P_x^{(1)} = P_y^{(1)}$,
then in 2-dimensional virtual coordinate system, $P_x^{(2)} =
P_y^{(2)}$. Since $P_x^{(1)} = P_y^{(1)}$
$$\Rightarrow P_x^{(1)} = (p_1 + d_x) = P_y^{(1)} = (p_1 + d_y)
\Rightarrow d_x = d_y$$

If $P_v^{(2)} = (p_1, p_2)$, then

$$\Rightarrow P_x^{(2)} = (p_1 + d_x, p_2 + d_x) = (p_1 + d_y, p_2 + d_y) = P_y^{(2)}$$

\end{proof}

\begin{lemma}
For any $k$-connected graph $G(V, E)$, the $D_d \geq k$.
\end{lemma}
\begin{proof}
We use contradiction to prove. Suppose $D_d < k$.
$$\Rightarrow \exists v_x, v_y \in V_c: P_x^{(D_d)} = P_y^{(D_d)}$$
We may simply elect $v_x$ to be a new dimension anchor, then
$$P_x^{(D_d + 1)} = (P_x^{(D_d)}, 0) $$
Since the network distance $\overrightarrow{v_xv_y}=d_{xy} > 0$
$$\Rightarrow P_y^{(D_d + 1)} = (P_y^{(D_d)}, d_{xy}) \neq P_x^{(D_d + 1)}$$
Contradict
\end{proof}

\begin{theorem}
The $D_d$ of a complete graph $G(V, E)$, is $|V|-1$.
\end{theorem}
\begin{proof}
A complete graph $G(V, E)$ is $|V|-1$-connected.
\end{proof}

\subsection{Greedy Forwarding Failure: Lack of Naming Uniqueness}

All routing failures of greedy forwarding on VCS including those in
the previous section and our previous work
\cite{paper:hgr,paper:avcs}, are caused by some nodes with the same
identity occurring in the network. For example, the {\em Expanded VC
Zone} anomaly and the {\em Disconnected VC Zone} anomaly in a
3-dimensional VCS \cite{paper:hgr} arise because the graph of the
network is 4-connected, which requires 4 or more anchors
(dimensions) to be present the network to produce a virtual
coordinate system with the maximal naming uniqueness. In some
randomly deployed network, the graph may be 1-connected or
2-connected. Anomalies in such network's VCS \cite{paper:avcs} is
caused by dimensional degradation, which may be only 1 or 2.
Increase the virtual coordinate dimensions does not increase the
naming uniqueness. In a word, the anomalies in VCS are caused by
either the limitation of dimensional degradation limiting the
uniqueness, or dimensions does not reach the dimensional degradation
-- its number of anchors is not sufficient.

Although based on the Papadimitriou-Ratajczak conjecture
\cite{paper:gfconjecture}, Rote and B\'{a}n\'{a}ny proved that a
greedy embedding exists in any given 3-connected planar graph
\cite{paper:greedydrawings,paper:greedydrawingsFull}. In reality, a
network with a 3-connected projected graph is not always available. A
typical deployed network contains some nodes with degree 1, and many
nodes with degree 2. Moreover, the required planarization may make it
even the situation worse by removing links, leading to a planar graph
with lower connectivity.  Thus, increasing the number of anchors
(dimensions) in VCS does not always make greedy forwarding always
successful due to degraded dimensionality.

Finally, for a complete graph, without a $|V|-1$-dimensional VCS,
there will always be some nodes with same identity (virtual
coordinates). But the $|V|-1$-dimensional VCS is no better than
shortest path routing which we want to avoid.

%% file: spvc.tex
\section{Spanning Path Virtual Coordinate System}
\label{sec:spvcs}

In section \ref{sec:vcdegrade}, we saw that the reason causing routing
failure in VCS is the lack of naming uniqueness.  In the worst case,
the connectivity of a network's mapping graph is 1. A consequent
observation is that a VCS constructed only by the network distance
(number of hops) to anchors can not provide naming uniqueness for
general graphs.  We propose here a new VCS naming approach, on which a
stateless routing protocol can guarantee packet delivery. We call the
VCS the Spanning-Path Virtual Coordinate System (SPVCS). In contrast
to existing VCS, SPVCS assumes a connectivity no bigger than 1.

\subsection{Spanning-Path VCS: Setup}

A good routing protocol must set up on a good naming base, which
should give each node an unique identity. Under the conservative
assumption that a network is only 1-connected, we can not depend on
increasing the number of anchors (or say, dimensions) of VCS as a way
to provide this uniqueness.

The design of SPVCS is based on a {\em depth-first search}
algorithm. A tree-style topology is constructed with only connection
information. Any node can be chosen as the naming root. Value 0 is
assigned to root node as its spanning-path virtual coordinate
(SPVC). The root node would start the naming process by sending a
{\em depth-first search naming} packet to one of its neighbor,
serving as its naming child. On receiving a {\em depth-first search
naming} packet, each node would be assigned an unique identity
(name) incrementally to the SPVC value of its sender. The sender of
this {\em depth-first search naming} packet is marked as the
receiver's parent. If a node has any neighbor that has not been assigned a
SPVC, it would send a {\em depth-first search naming} packet to this
neighbor. Otherwise, if all its neighbors are assigned a SPVC
accordingly, it would decide that it is an end on the
spanning-path, and sends an {\em end-of-search naming} packet to its
parent, containing the maximal SPVC value of all its children. On
receiving an {\em end-of-search naming} packet, a node would either
send another {\em depth-first search naming} packet to one of its
neighbors which has not been assigned a SPVC, with the replied SPVC
value in the received {\em end-of-search naming} packet if
applicable. Or if all its neighbors are assigned with SPVC, it would
forward this {\em end-of-search naming} packet to its parent. This
process would repeat until the root node receives an {\em
end-of-search naming} packet and finds all of its neighbors are
assigned some SPVC.  As long as the network is connected, all nodes receive a unique identifier.

The pseudo-algorithm of SPVC naming process is summarized algorithm
\ref{algo:spvcs}, where the $SetupSPVC(node)$ is a recursive function
used to set the spanning-path virtual coordinate values of $node$, as
algorithm \ref{algo:spvc}.

\begin{algorithm}\small
\caption{Set up the SPVCS of network} \label{algo:spvcs}
\begin{algorithmic}
\STATE anchor $\leftarrow$ $AnchorSelectFunction()$

\STATE anchor.spvc $\leftarrow$ 0

\STATE anchor.max\_range $\leftarrow$ 0

\STATE anchor.max\_range $\leftarrow$ $SetupSPVC(anchor)$
\end{algorithmic}
\end{algorithm}

\begin{algorithm}\small
\caption{Set up Spanning-Path Virtual Coordinate Recursively}
\label{algo:spvc}
\begin{algorithmic}
\STATE {\bf Function} $SetupSPVC(root)$
    \FORALL{node in root.neighbors}
        \IF{node is not set}
        \STATE node.parent $\leftarrow$ root
        \STATE node.spvc $\leftarrow$ root.max\_range + 1
        \STATE node.max\_range $\leftarrow$ node.spvc
        \STATE root.max\_range $\leftarrow$ $SetupSPVC(node)$
        \ENDIF
    \ENDFOR
    \STATE {\bf Return} $root.max\_range$
\end{algorithmic}
\end{algorithm}

\begin{figure}[t]
\centering
\includegraphics[width=0.35\textwidth]{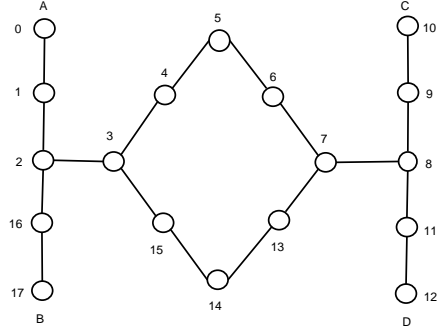}
\caption{Example: Spanning-Path VCS} \label{fig:spvcsExample}
\end{figure}

\subsection{Spanning-Path Geometric Routing}
Based on the {\em Spanning-Path Virtual Coordinate System} (SPVCS),
a stateless and delivery guaranteed geometric routing can be
constructed. Suppose a node with SPVC value $x$ (referred as node X)
needs to send a packet to another node with SPVC value $y$ (referred
as node Y). It would mark each neighbor's {\em range} as the
neighbor's  SPVC value and its {\em max-child} SPVC value. A
neighbor whose range contains the destination node's SPVC value $y$,
is called a {\em forwarding candidate}. There are at most two
forwarding candidates among the node's neighbors, one of which is
its parent. The non-parent forwarding candidate is preferred. This
process would be repeated by any node receiving this packet, until
node Y receives the packet. The algorithm can be summarized as
algorithm \ref{algo:spr}.

\begin{algorithm}\small
\caption{Spanning Path Routing on VCS} \label{algo:spr}
\begin{algorithmic}
\STATE {\bf Procedure} $SPR(source, destination)$
    \IF{source $=$ destination}
    \STATE {\bf Return}
    \ENDIF
    \FORALL{node in source.neighbors}
        \IF{node $\ne$ source.parent}
            \STATE range $\leftarrow$ (node.spvc, node.max\_range)
            \IF{destination.spvc $\in$ range}
                \STATE nexthop $\leftarrow$ node
                \STATE $SPR(nexthop, destination)$
                \STATE {\bf Return}
            \ENDIF
        \ENDIF
    \ENDFOR
\STATE nexthop $\leftarrow$ source.parent
\STATE $SPR(nexthop,
destination)$
\end{algorithmic}
\end{algorithm}

{\bf Spanning-Path Routing Example} Let's use the SPVC value of each
node as its ID since this value is unique. In the figure
\ref{fig:spvcsExample}, the node 5 needs to send a packet to node
14. In the VCS shown in figure \ref{fig:dimDegradation}, the greedy
forwarding will fail since the source has the same identity as
destination. In SPVCS, node 5 first check all its neighbors' ranges:
node 4's range is (4, 15), node 6's range is (6, 15). the
destination SPVC is 14, so both neighbors are forwarding candidates.
And node 4 is node 5's parent, node 5 would forward the packet to
node 6. Node 6 will repeat the process to forward packet to node 7.
Node 7's neighbors range are node 6 (6, 15), node 8 (8, 12) and node
13 (13, 15). Node 7 would forward packet to node 13. And finally,
node 13 forwards packet to node 14.

As we can see, the spanning-path geometric routing is stateless and
definitive: any forwarding node only needs the SPVCs of all its
neighbors and the destination to make routing decision, without any
repeat link on path.

\begin{theorem}
Spanning-Path Geometric Routing is delivery guaranteed, if the
network is connected.
\end{theorem}
\begin{proof}
Since $G(V,E)$ is connected we have
$$\forall v \in V: v.spvc \in anchor.range$$
$$\forall v \in V: v.spvc \in v.parent.range$$
$$\Rightarrow \text{anchor is definitive reachable}$$
where {\em definitive reachable} means no repeat link on path. And
$$\forall v \in V: \exists n \in \text{anchor.neighbors}, v.spvc \in
n.range$$
$$\Rightarrow v \text{ is definitive reachable}$$

\end{proof}

\subsection{Optimized Spanning-Path VCS}
\begin{figure}[t]
\centering
\includegraphics[width=0.35\textwidth]{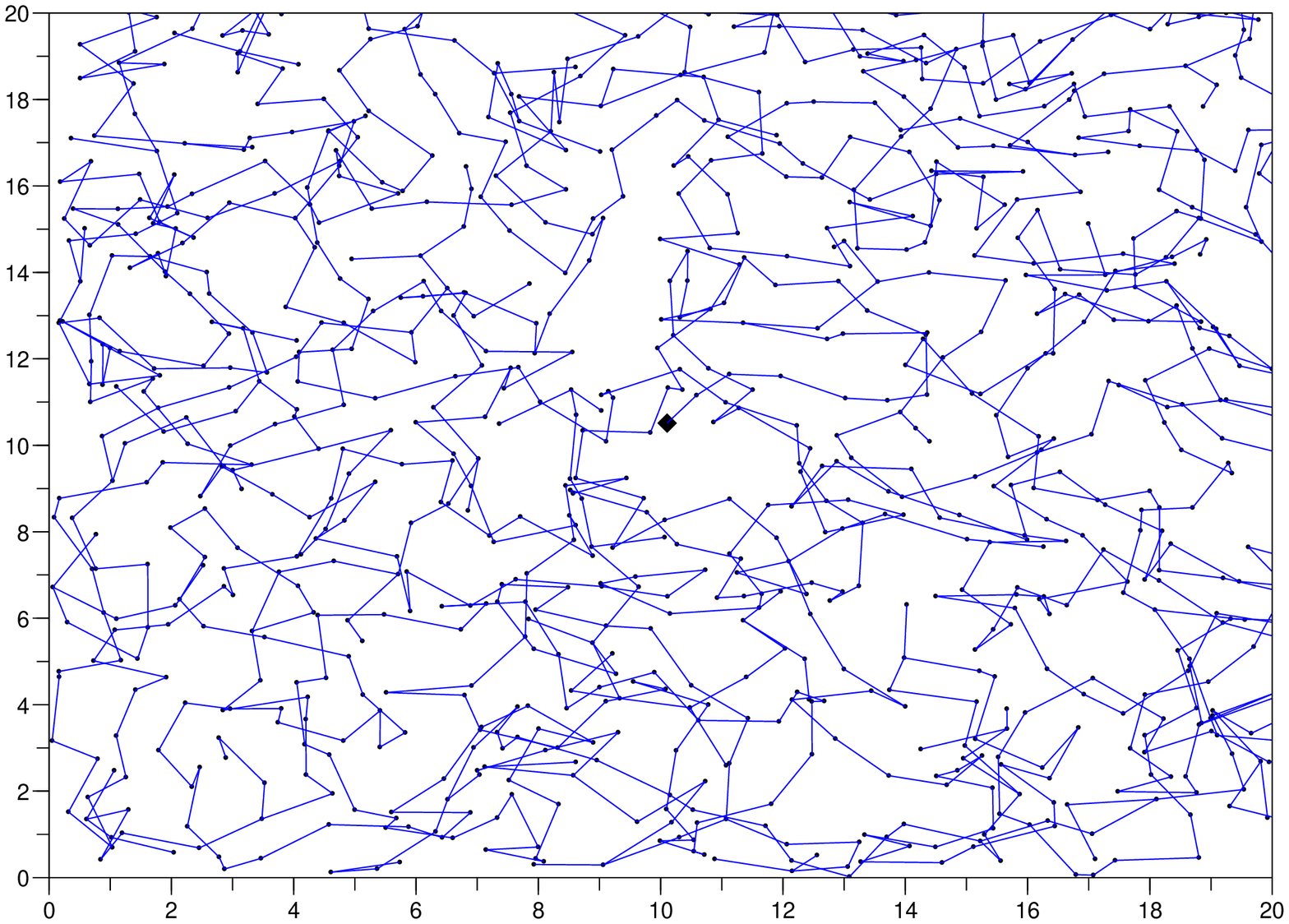}
\caption{Sample SPVCS: Anchor at Center}
\label{fig:sample-spvcs-centeranchor}
\end{figure}

\begin{figure}[t]
\centering
\includegraphics[width=0.35\textwidth]{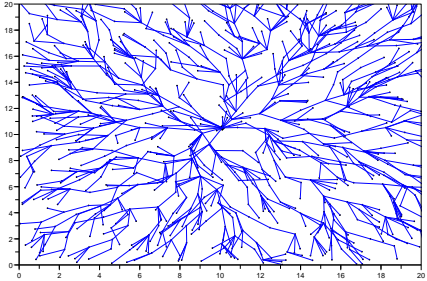}
\caption{Sample: Optimized SPVCS -- Anchor at Center}
\label{fig:optimized-spvcs-centeranchor}
\end{figure}

The DFS based constructing procedure of Spanning-Path VCS leads to
an un-balanced tree, shown as figure
\ref{fig:sample-spvcs-centeranchor}. Quite pathes constructed on
Spanning-Path VCS need to go through the anchor node. A constructing
procedure based on the {\em breadth-first search} (BFS) lead SPVCS
to a balanced tree topology, shown in figure
\ref{fig:optimized-spvcs-centeranchor}. The algorithm can be
summarized as algorithm \ref{algo:ospvcs},
\begin{algorithm}\small
\caption{Set up the parent of nodes in Optimized Spanning-Path VCS}
\label{algo:ospvcs}
\begin{algorithmic}
\STATE {\bf Procedure} $SetupParent(anchor)$
    \STATE anchor.parent $\leftarrow$ anchor
    \STATE enqueue(anchor)
    \WHILE{ queue is not empty}
        \STATE node $\leftarrow$ dequeue()
        \FORALL{n in node.neighbors}
            \IF{n.parent is not set $and$ n is not in queue}
                \STATE n.parent $\leftarrow$ node
                \STATE enqueue(n)
            \ENDIF
        \ENDFOR
    \ENDWHILE

    \STATE anchor.spvc $\leftarrow$ 0
    \STATE anchor.max\_range $\leftarrow$ 0
    \STATE anchor.max\_range $\leftarrow$ $SetupOSPVC(anchor)$
\end{algorithmic}
\end{algorithm}
where the $SetupOSPVC(node)$ is shown as algorithm \ref{algo:ospvc}.
\begin{algorithm}\small
\caption{Set up the Optimized Spanning-Path Virtual Coordinate}
\label{algo:ospvc}
\begin{algorithmic}
\STATE {\bf Function} $SetupOSPVC(root)$
    \FORALL{n in root.neighbors}
        \IF{n.parent is root}
            \STATE n.spvc $\leftarrow$ root.max\_range
            \STATE n.max\_range $\leftarrow$ n.spvc
            \STATE root.max\_range $\leftarrow$ $SetupOSPVC(n)$
        \ENDIF
    \ENDFOR
    \STATE {\bf Return} root.max\_range
\end{algorithmic}
\end{algorithm}

\subsection{Aligned Greedy and Spanning-Path Routing (AGSP)}
Since the spanning-path routing does not provide a greedy algorithm
which shows a performance comparable to the optimal solution -- the
shortest path routing \cite{paper:avcs}, to use
SPR as complementary routing to greedy forwarding is rational. As we
will show with experiment in section \ref{sec:experiment}, SPR
collaborating with greedy forwarding will generate path with much
better stretch. The routing algorithm of aligned greedy and
spanning-path routing can be summarized as algorithm
\ref{algo:agsp},
\begin{algorithm}\small
\caption{Aligned Greedy and Spanning-Path Routing} \label{algo:agsp}
\begin{algorithmic}
\STATE $SetupAVCS(Network)$

\STATE $SetupSPVCS(Anchor)$

\FORALL{node src in Network}
    \FORALL{node dst in Network - \{src\}}
        \STATE current $\leftarrow$ src
        \WHILE{current $\ne$ dst}
            \STATE nexthop $\leftarrow$ $GFonAVCS(nexthop, dst)$
            \IF{nexthop $=$ current}
                \STATE $SPR(current, dst)$
            \ELSE
                \STATE current $\leftarrow$ nexthop
            \ENDIF
        \ENDWHILE
    \ENDFOR
\ENDFOR
\end{algorithmic}
\end{algorithm}
where the $SetupAVCS(Network)$ is to set up the aligned virtual
coordinate system, on which the greedy forwarding can be used as
$GFonAVCS(src, dst)$.

%% file: experiment.tex
\section{Experimental Evaluation}
\label{sec:experiment}

In this section, we present an experimental evaluation of the Spanning
Path Virtual Coordinate System (SPVCS) in comparison to the
complementary routing protocols in existing geometric protocols on
physical coordinates (GeoCS) and virtual coordinates Systems (VCS).
The evaluation tracks the average path stretch relative to SP. We also
simulate AGSP, which uses aligned virtual coordinates, and switches to
SPVCS when anomalies are encountered.  We use a custom simulator
written in C, to abstract away the details of the channel and
networking protocols.

We study both random (uniform) and a custom ``C'' deployment.  In
the uniform scenarios, each node's location is generated uniformly
across the simulation area.  For these scenarios, each point
represents the average of 30 scenarios of 1000 nodes that are
deployed $100\times100$ unit area; in the custom scenarios, each
points represents the average of 30 scenarios of 150 nodes deployed
in a "C" style area to create a physical void.  In both cases, the
number of scenarios was sufficient to tightly bound the confidence
intervals. We simulate different densities by varying the radio
transmission range. For every scenario, reach-ability is determined
by testing packet delivery success between each pair of nodes in the
network. Recall that the stateful SP provides optimal routing in
terms of number of hops; for this reason it is used as the baseline
for ideal performance in terms of path stretch.

\subsection{Spanning-Path vs Perimeter Routing}

\begin{figure}[t]
\centering
\includegraphics[width=0.38\textwidth]{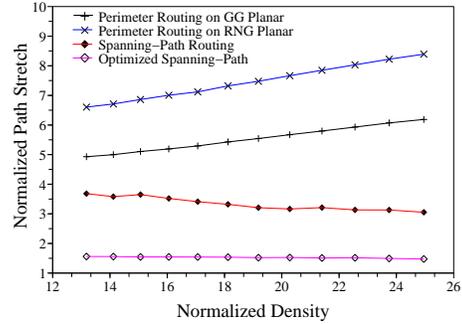}
\caption{Path Stretch: Perimeter Routing vs Spanning-Path Routing}
\label{fig:complementaryPathStretch}
\end{figure}

Figure \ref{fig:complementaryPathStretch} shows the path stretch of
Perimeter routing (which is the complimentary routing protocol in
GPSR) with different planarization algorithms.  The figure also shows
the performance of Spanning-Path Routing on SPVCS. As the density goes
higher, Perimeter routing suffers; this is a known problem for
perimeter routing, leading to an increased path stretch.  However,
Spanning-Path routing benefits from the denser network because it is
based on connectivity instead of physical distance.

\begin{figure}[t]
\centering
\includegraphics[width=0.3\textwidth]{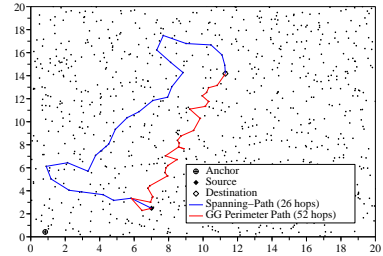}
\caption{Sample Path: Spanning-Path vs Perimeter}
\label{fig:samplepath}
\end{figure}

Figure \ref{fig:samplepath} shows a comparison of sample paths
constructed by Perimeter routing on GG Planar graph and Spanning-Path
routing. Although the path constructed by Spanning-path routing is
longer in distance length, it is much shorter in number of hops.

\subsection{Aligned Greedy Spanning-Path routing}
\begin{figure}[t]
\centering
\includegraphics[width=0.37\textwidth]{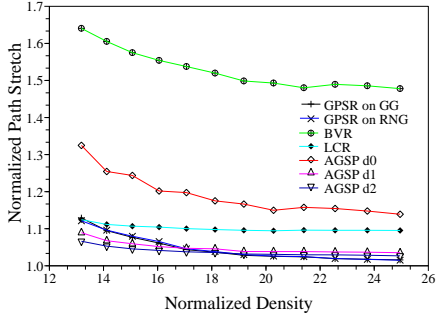}
\caption{Path Stretch: Geometric Routings}
\label{fig:GeometricRoutingPathStretch}
\end{figure}
Figure \ref{fig:GeometricRoutingPathStretch} shows the path stretch
obtained by different geometric routing protocols. As we can see, with
some VCS alignment, AGSP (Aligned Greedy Spanning-Path) routing
provides a competitive performance to that of GPSR, especially in
sparse scenarios. Without alignment, greedy spanning-path routing
suffers from the low greedy ratio due to the 4-d VCS naming failures.

\subsection{Custom Deployment}

\begin{figure}[t]
\centering
\includegraphics[width=0.37\textwidth]{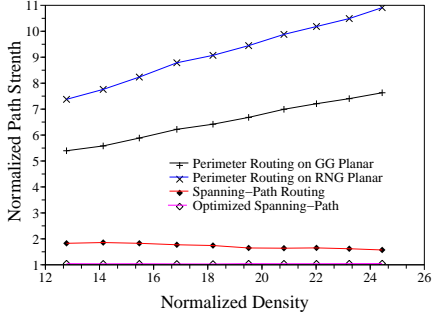}
\caption{Complementary Routing on C Topology}
\label{fig:cvoid-complementary-pathstretch}
\end{figure}

To study the protocols under more demanding conditions when uniform
coverage does not exist, nodes are deployed uniformly in a C pattern,
leaving a significant void area.  Figure
\ref{fig:cvoid-complementary-pathstretch} shows the path stretch of
Perimeter and Spanning-path routings in such scenarios scenario. As we
can see, Spanning-path can tolerate such scenarios, while Perimeter
routing suffers poor performance.

Also, AGSP shows a nearly optimal path stretch in such scenario
compared to shortest path routing. Meanwhile, GPSR performance suffers
since its greedy ratio drops dramatically due to the presence of voids
on many paths. LCR shows a similar performance because its
backtracking algorithm uses blind search of a limited number of neighbors.
\begin{figure}[t]
\centering
\includegraphics[width=0.37\textwidth]{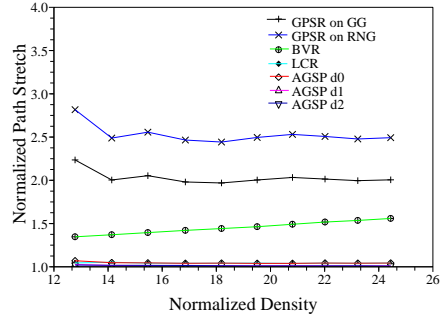}
\caption{Path Stretch: C Topology}
\label{fig:cvoid-gr-pathstretch}
\end{figure}

\subsection{The Impact of Anchor Location on SPVCS}

\begin{figure}[t]
\centering
\includegraphics[width=0.35\textwidth]{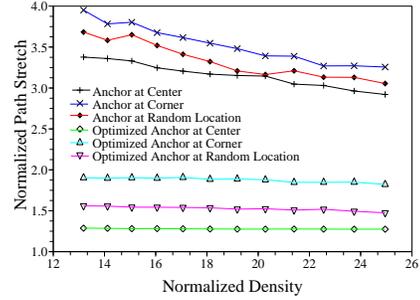}
\caption{Path Stretch: Impact of Anchor's Location}
\label{fig:anchor-impact}
\end{figure}

Intuitively, anchor location has significant impact on the performance
SPVCS.  Experimental results support this intuition (Figure
\ref{fig:anchor-impact}). An anchor node located near the center leads
to the best path stretch since it can provide a more balanced
spanning-tree.  Conversely, an anchor at the corner results in worse
performance with respect to path stretch.


\subsection{Optimizing SPVCS}

\begin{figure}[t]
\centering
\includegraphics[width=0.3\textwidth]{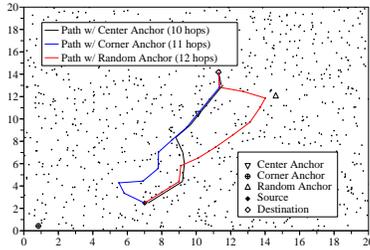}
\caption{Sample Path: Optimized SPVCS}
\label{fig:optimized-spvcs-samplepath}
\end{figure}

Figure \ref{fig:sample-spvcs-centeranchor} shows a sample SPVCS. The
SPVCS algorithm leads to an unbalanced spanning tree of the network;
clearly this is not the most efficient spanning topology.  We seek to
optimize this topology by replacing it with a balanced spanning tree,
creating more effective paths, and limiting the impact of the anchor's
location. Sample paths are shown in
Figure~\ref{fig:optimized-spvcs-samplepath}.

%
%
%
%
%

%% file: conclusion.tex
\section{Conclusion}
\label{sec:conclusion}

In this paper, we first analyze the reasons behind geometric routing
failure in recently proposed VCS: dimensional degradation leading to
the lack of uniqueness in naming. Practically, a unique identity can
not be easily assigned to any node in VCS on network, due to
limitation of network connectivity. This analysis represents a
contrast to the common assumption of most those virtual coordinate
systems -- the more anchor nodes (virtual coordinate dimensions), the
better the uniqueness and routability.

Consequently, we propose an alternative naming algorithm for virtual
coordinate systems for geometric routing protocols, in which only one
dimension (anchor) is used. We call this naming algorithm the
Spanning-path Virtual Coordinate System. SPVCS provides unique numbers
to all nodes in the network in a way that allows greedy routability
(albeit with some path stretch since SPVCS does not use the full
connectivity information).

Upon this SPVC assignment, a stateless and delivery guaranteed
geometric routing protocol is constructed. We show that this protocol
outperforms geographic routing (GPSR) \cite{paper:gpsr}, and several
recently proposed geometric routing protocols on virtual coordinate
systems such as LCR \cite{paper:lcr}, BVR \cite{paper:bvr} and AVCS
\cite{paper:avcs}.